\documentclass{emulateapj}
\def\figdir{.}

\usepackage{graphicx}
\usepackage{txfonts}

\slugcomment{Submitted to ApJ}
\shorttitle{MHD simulations of penumbra fine structure}
\shortauthors{Heinemann et al.}

\newcommand{\AN}{\AA N}

\begin{document}
\title{MHD simulations of penumbra fine structure}

\author{T. Heinemann\altaffilmark{1}}
\affil{Department of Applied Mathematics and Theoretical Physics,
  Centre for Mathematical Sciences, Wilberforce Road, Cambridge CB3
  0WA, United Kingdom} 
\author{\AA. Nordlund}
\affil{Niels Bohr Institute, University of Copenhagen, Juliane Maries
  Vej 30, 2100 Copenhagen, Denmark}
\author{G.B Scharmer}
\affil{Institute for Solar Physics, Royal Swedish Academy of Sciences,
  AlbaNova University Center, SE-106\,91 Stockholm, Sweden}
\and
\author{H.C. Spruit}
\affil{Max Planck Institute for Astrophysics, Box 1317, 85741
  Garching, Germany} 

\altaffiltext{1}{Previously at:  NORDITA, AlbaNova University Center,
  10691 Stockholm, Sweden} 
                                %

\begin{abstract}
  We present results of
  numerical 3D MHD simulations with radiative energy transfer of fine
  structure in a small sunspot of about 4~Mm width. The simulations
  show the development of filamentary structures and flow patterns
  that are, except for the lengths of the filaments, very similar to
  those observed. The filamentary structures consist of gaps with
  reduced field strength relative to their surroundings. Calculated
  synthetic images show dark cores like those seen in the
  observations; the dark cores are the result of a locally elevated
  $\tau=1$ surface. The magnetic field in these cores is weaker and
  more horizontal than for adjacent brighter structures, and the cores
  support a systematic outflow. Movies show migration of the
  dark-cored structures towards the umbra, and fragments of magnetic
  flux that are carried away from the spot by a large scale `moat
  flow'. We conclude that the simulations are in qualitative agreement
  with observed penumbra filamentary structures, Evershed flows and
  moving magnetic features.

\end{abstract}

\keywords{sunspots -- magnetic field}

\section{Introduction}

Explaining the magnetic field, Evershed flow, overall fine structure
of sunspot penumbrae and the observed high heat flux ($\sim$75\% of
that of the quiet sun atmosphere), constitute some of the most
longstanding problems in solar physics. Despite nearly a century of
detailed studies, no consensus has been reached as regards the origin,
or even morphology, of this fine structure. A major obstacle to
progress is our inability to adequately resolve sunspot fine structure
when observations are carried out with spectrographs or
spectropolarimeters. The discovery of dark-cored penumbra filaments
(Scharmer et al.\@ 2002) suggests that the basic elements of penumbral
structure are now observable with imaging instrumentation.

Ambiguities associated with interpretations of existing observational
data have given rise to a diversity of models and explanations for
flows and fine structure in penumbrae (e.g., Meyer \& Schmidt 1968,
Solanki \& Montavon 1993, Schlichenmaier et al.\@ 1998a,b, Thomas et
al.\@ 2002, Spruit and Scharmer 2006). In contrast to what is
the case for granulation (e.g., Stein and Nordlund 1998), faculae
(Carlsson et al.\@ 2004, Keller et al.\@ 2004, Steiner 2005) and
umbral dots (Sch\"ussler \& V\"ogler 2006), realistic numerical
simulations of entire sunspots have not yet been feasible. This is
partly due to difficulties of thermally relaxing such a deep structure
and maintaining its stability, but mostly due to the huge range of
scales associated with a fully developed sunspot.

\subsection{The gappy penumbra}

Spruit and Scharmer (2006 -- hereinafter SS06a) have pointed to a
similarity between penumbral filaments and light bridges, the bright
lanes often seen crossing an umbra. Both have a dark core: a narrow
dark lane along their mid-line. In SS06a it was argued that this is
not just an observational congruence, but that the similarity actually
indicates a common origin for these structures: gaps in the magnetic
field, closing near the visible surface, and heated by convection in
the gap. In Scharmer and Spruit (2006 -- hereinafter SS06b) magnetostatic
models for such structures were presented. Their properties,
distinctly different in the inner and outer penumbra, are
  consistent with a number of observations, such as the differences
in strength and inclination of the magnetic field between dark and
bright penumbral structure.

Radiative MHD simulations of light bridges by one of the authors (\AN,
unpublished) and Heinemann (2006) show how the dark cores over the
center of the light bridges are formed. They are found to be a
consequence of hydrostatic pressure and radiative energy balance
around the `cusp' of the magnetic field, at the height in the
atmosphere where field lines from both sides close over the gap.
A second
observational clue is given by umbral dots, which are often seen to
evolve from the `heads' of penumbral filaments propagating into the
umbra. As proposed by Parker (1979), seen in numerical simulations by
Nordlund \& Stein (1990), and recently confirmed through high
resolution numerical simulations by Sch\"ussler and V\"ogler (2006),
umbral dots are the surface manifestation of nearly field-free
  gaps below the umbral photosphere. This theoretical evidence is
strengthened observationally by the short dark cores seen to cross
peripheral umbral dots (Scharmer et al.\@ 2002).

High-resolution images of complex sunspot groups (e.g., Fig.\@~1 of
Rouppe van der Voort et al.\@ 2004) show penumbra structures associated
with both small and large, symmetric and irregular sunspots, and even
penumbrae without visible umbrae. This suggests that there is no need
to model entire sunspots to explain penumbra fine structure. This, and
the rapid formation of penumbrae with well developed Evershed flows
(Leka \& Skumanich 1998, Yang et al.\@ 2003), also suggest that details
of the deep structure of the magnetic field may not be crucial for
these structures. In this paper, we describe first results from
realistic simulations of penumbral fine structure. The model used for
these simulations is a rectangular section of the solar atmosphere
containing a part of a sunspot with umbra and penumbra, embedded in an
essentially field-free atmosphere. The simulations show the
development of short filamentary structures, associated with the
formation of gaps {with strongly reduced field strength}. They
resemble observed penumbral filaments as seen in the inner penumbra:
like these, they have a characteristic head-and tail structure with a
dark core running along the tail, and are seen to propagate inward.

\section{MHD Simulations}

The simulations were carried out using the PENCIL code\footnote{see
  http://nordita.dk/software/pencil-code}, modified to handle energy
transfer by radiation in a grey atmosphere (Heinemann 2006, Heinemann
et al.\@ 2006). We used a rectangular computational box
$12448\times6212$~km in the horizontal ($x$ and $y$)
directions, extending over a depth ($z$) range of $3094$~km and
with a grid separation of $24.36$~km in both horizontal and vertical
directions ($512\times256\times128$ grid points). The quiet sun
photosphere is located approximately $700$~km below the upper
boundary. In the horizontal directions, periodic boundary conditions
are used. Both the upper and lower boundaries are closed in the sense
of not allowing vertical flows. The vertical magnetic field component
is kept fixed at the lower boundary. To avoid inconsistencies with that boundary
condition, the horizontal velocity is set to zero at the lower boundary. 
The magnetic field is free to evolve at the upper boundary.
 
The initial magnetic field of the simulations is a potential
(current-free) magnetic field, initially depending only on $x$
  and $z$, separated by a current sheet from the surrounding
convection zone.  This initial magnetic field is defined by two free
parameters, the total vertical flux and the ratio of gas pressure to
magnetic pressure, $\beta$, along the current sheet. For all
calculations made, $\beta$ was assumed to be independent of depth. The
corresponding potential field, given by magnetostatic equilibrium with
a gas pressure obtained from a convection simulation and the chosen
value of $\beta$, was calculated with the method described in SS06b.
Since the subsequent magnetic field evolution is driven by the
internal dynamics of the model, the choice of the total vertical flux
and $\beta$ primarily constrains the width and field strength of the
flux bundle close to the lower boundary.

\begin{figure}[tbh]
  \includegraphics[width=\linewidth]{\figdir/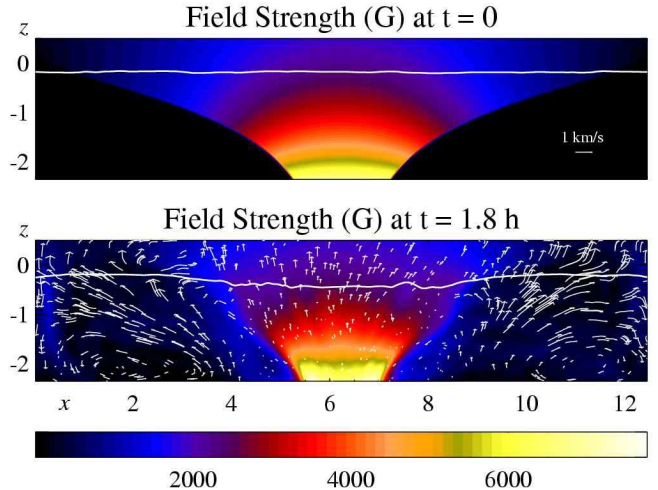}
  \caption{\small Magnetic field strength for the initial state
    (upper panel) and averaged over the y-axis for the snapshot
      seen in Fig.\@~2 (lower panel). Also shown are contours of 
    constant optical depth $\tau = 1$. The lower panel shows the
    average flow field (arrows).}
  \label{cores1}
\end{figure}
We made several simulations with different values for the total
vertical flux and $\beta$. All simulations showed the formation of
dark-cored filamentary structures, but with variations in their
morphology, number density and life times. Here we discuss only the
results of one of these simulations, obtained with an average vertical
field strength of 1~kG and $\beta=19$, corresponding to an initial
ratio of the gas pressures in the magnetic to non-magnetic
  components of $\beta/(\beta + 1)=0.95$. The initial magnetic field
configuration is shown in Fig.\@~1 (upper panel).

The temperature, gas pressure, and velocity field were initialized
from a field-free convection simulation. The gas pressure in the
magnetic atmosphere was then reduced by multiplication of the mass
density by  $\beta/(\beta + 1)$ and the magnetic field was added to
the model. The strong suppression of convection by the magnetic field
causes a rapid drop in temperature, leading to a strong reduction in
gas pressure above the umbra and hence a partial collapse of the
magnetic field configuration. After approximately 30 solar minutes,
the gas reached a quasi-equilibrium state. The magnetic field strength
and flow field (arrows) after approximately 1.8 solar hours, averaged
over the y-direction, are shown in Fig.\@~1 (lower panel). 

We assumed a 6th order hyperdiffusivity of $2\cdot10^{-9}$ and a shock
diffusivity of 5 (cf.\@ Eqs.\@~(130) and~(44) of the PENCIL code
documentation -- see the footnote), in units where length is measured
in megameters and time is measured in kiloseconds. The
hyperdiffusivity is adequate to avoid the development of ripples at
the grid scale, while leaving larger structures essentially inviscid
and non-resistive. The shock diffusivity ensures that shocks -- which
are ubiquitous in the upper photosphere -- are adequately resolved by
the difference scheme. Radiative energy transfer was handled using 6
rays along the coordinate axes. 

\section{Results}
\subsection{Dark-cored filamentary structures}

Following the initial dramatic adjustment of the model to the applied
magnetic field, convection forms small intrusions in the magnetic atmosphere, propagating towards the umbra. Dark lanes, similar to the dark cores of penumbral
filaments (Scharmer et al.\@ 2002), form rapidly. A statistically
stable situation is reached, where filaments continuously form and
disappear on time scales on the order of 30~min.
\begin{figure}[t]
  \includegraphics[width=1.00\linewidth]{\figdir/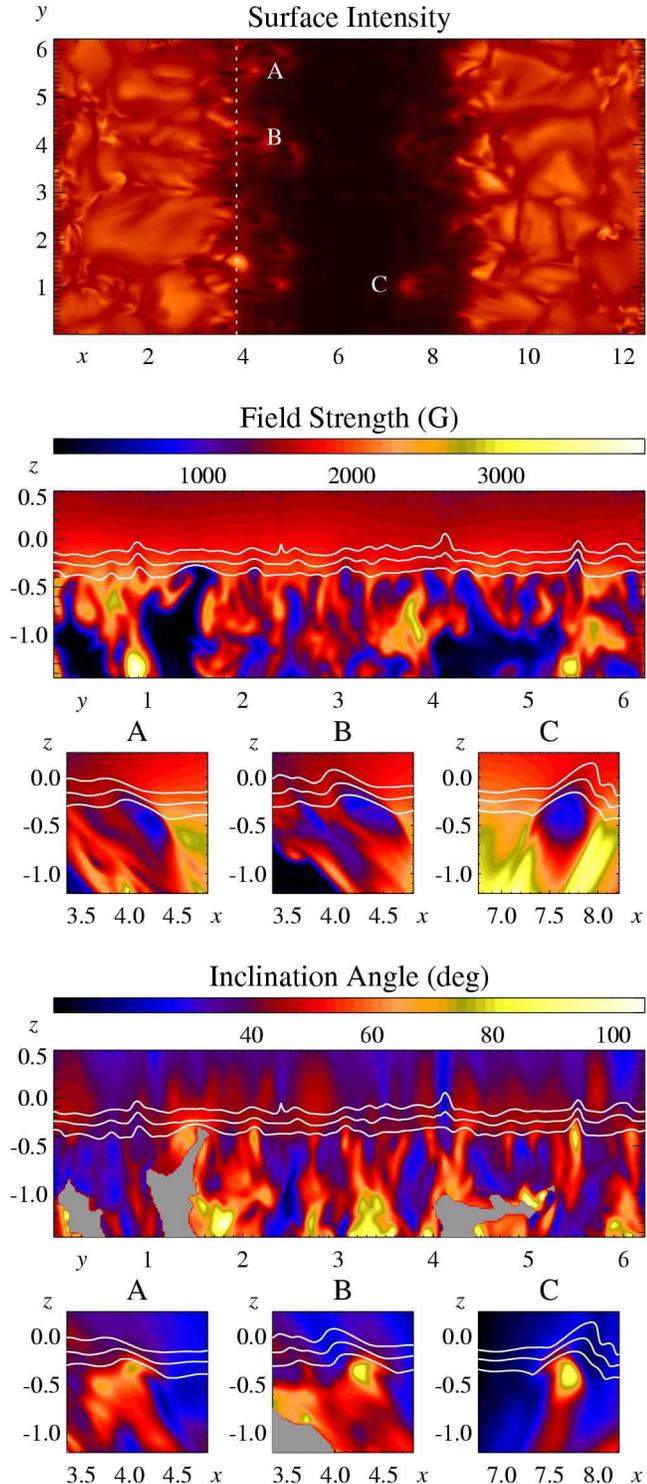}
  \caption{\small Emergent intensity and false-color images of the
    magnetic field strength and inclination along the dotted line in
    the intensity image.  The six small panels show the field strength
    and inclination for cuts along three dark-cored filamentary
    structures (indicated A--C) in the direction perpendicular to the
    dotted line. The white curves indicate the heights for which 
    (top to bottom) $\tau = 0.01, 0.1, 1$. Tickmarks and axis labels 
    indicate units in Mm.}
  \label{cores2}
\end{figure}

Figure 2 shows a snapshot after 1.8 solar hours of the emergent
intensity and the field strength for a vertical slice of the model at
the location of the dotted line in the intensity image. The
intensity image shows several short dark-cored filamentary structures
protruding into the umbra. The cross-section of the field strength
shows several gaps with strongly reduced field strength near the
visible surface, while deeper down the magnetic field is
strongly fragmented with many and nearly field-free gaps. Also 
shown in Fig.\@~2 is the magnetic field strength and inclination of
three dark-cored filaments along cuts perpendicular to the dotted line
in the intensity image. These cuts show intrusions with strongly
  reduced field strength that extend over approximately 400~km in
depth, and that appear similar to the nearly field-free gaps in
the umbral dot simulations of Sch\"ussler and V\"ogler (2006). These
cuts also demonstrate that the gaps are tilted and roughly aligned
with the surrounding magnetic field and that the field is nearly
  horizontal at locations of these filamentary structures near the
  surface.  Shown in Fig.\@~2 are also contours of constant optical
depth for $\tau = 0.01$, 0.1, and 1 (top to bottom). Three of the
dark-cored filaments are associated with $\tau=1$ surfaces that are
locally elevated by approximately 100--200~km, implying a strongly
localized warping of the visible surface. At heights of only 100--200~km above
the $\tau=1$ layer, the magnetic field is largely homogeneous, with
only small horizontal variations in field strength.

  In Fig.\@~3 is shown horizontal variations in the field strength
  and inclination angle at the height where (locally) $\tau=1$ for
  the left-hand side of the umbra-penumbra shown in Fig.\@~2. This
  shows several filamentary structures with strongly reduced field
  strength and nearly horizontal magnetic field at the center of the
  filaments. Several of these have widths that are only slightly in
  excess of 100~km, suggesting that they are not adequately resolved
  with the grid separation used. Figure 3 also shows that the gaps are
  more numerous, but otherwise similar, 200~km below the visible surface.
\begin{figure}[htbp]
  \includegraphics[width=1.00\hsize]{\figdir/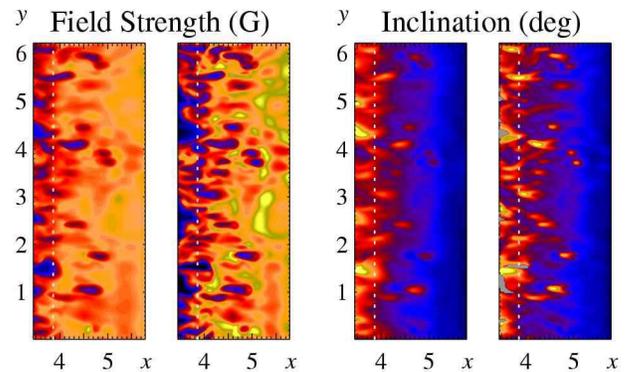}
  \caption{\small False-color images of the field strength and
      inclination angle at $\tau=1$  (left) and 200~km below $\tau=1$ (right) for the left part
      of the umbra-penumbra shown in Fig.\@~2. The color look-up tables are the same as shown
      in Fig. 2}
  \label{cores3}
\end{figure}

\begin{figure}[htbp]
  \includegraphics[width=1.00\hsize]{\figdir/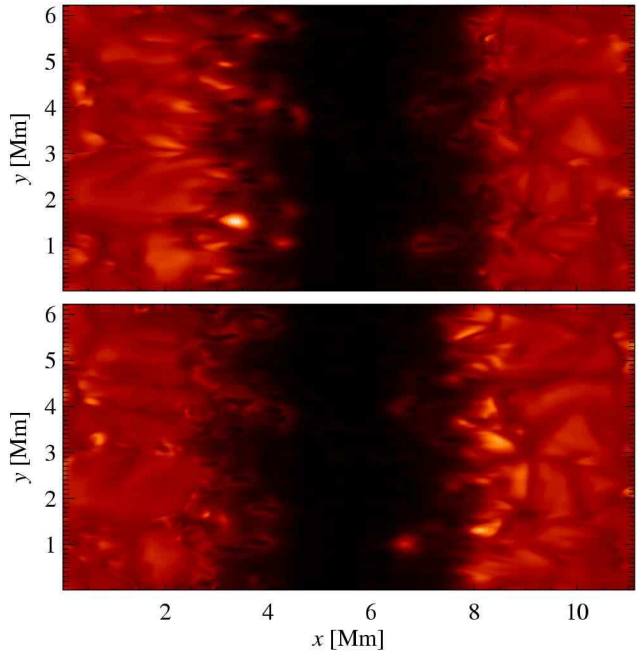}
  \caption{\small Calculated synthetic images corresponding to a
    heliocentric distance of 45$^\circ$. In the top image, the limb
    direction is left, in the bottom image the limb direction is
    right. To facilitate comparison with the corresponding disk-center
    image (Fig.~2), the fore-shortening of the image in the horizontal
    direction has been compensated for.}
  \label{cores4}
\end{figure}

\begin{figure}[tbh]
  \includegraphics[width=1.00\hsize] {\figdir/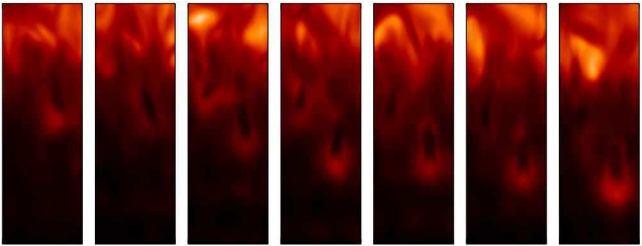}
  \caption{\small Evolution and inward migration of the dark-cored
    structure labeled `C' in Fig.\@~2 during 20 solar minutes. The
    last snapshot is from the larger field-of-view shown in Figs.\@~2
    and~3.} 
  \label{cores5}
\end{figure}

Fig.\@~4 shows the calculated emergent intensities corresponding
to a heliocentric distance of $45^\circ$. On the limb-side, at least 
five round umbral dot-like structures are seen. On the disk-center side, 
the same dots are fainter, less distinct and partly masked by dark cores.
These differences between the limb side and disk center side of
penumbrae are in excellent qualitative agreement with observations
(Tritschler et al. 2004, S\"utterlin et al.\@ 2004, Langhans et al.\@ 2006).
The explanation for the enhanced visibility of bright dots on the limb side is
essentially the same `limb brightening' effect seen in faculae: the
line of sight is more perpendicular to the $\tau=1$ surface on the
limb side, allowing us to see deeper into the hotter layers of the
convecting gas (Spruit 1976, Carlsson et al.\@ 2004, Keller et al.\@
2004).

Movies of the synthetic intensity images show dark-cored filamentary
structures forming and migrating toward or into the umbra.  Several of
these structures are significantly darker than observed
penumbral filaments, resembling instead the faint dark-cored
structures described by Scharmer et al.\@ (2002).  Figure 5 shows
snapshots taken at 86.7, 90, 93.3, 96.7, 100, 103.3, and 106.7 solar
minutes for one such faint dark-cored filament (labeled `C' in
Fig.\@~2), demonstrating the development of a dark core and the inward
migration of the structure. During the time interval of 20 solar
minutes shown in this figure, this structure propagates approximately
850~km toward and into the umbra, corresponding to a propagation
speed of 0.7~km\,s$^{-1}$. This is similar to the inward propagation
speeds of approximately 0.5--1~km\,s$^{-1}$ found by Rimmele and Marino
(2006) and the 300~m\,s$^{-1}$ found by Langhans et al.\@ (2006). 
\subsection{Flows associated with dark-cored filaments}
\begin{figure}[t]
  \includegraphics[width=1.00\hsize] {\figdir/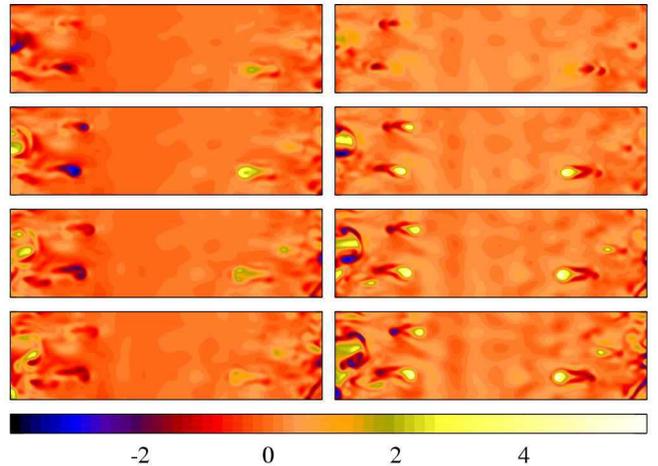}
  \caption{\small Horizontal (left) and vertical velocities at
    $z=-0.23$, $-0.35$, $-0.47$, and $-0.60$ (top to bottom) for the
    dark-cored structures on opposite sides of the umbra in the lower
    part of Fig.\@~2.}
  \label{cores6}
\end{figure}
Figure 6 shows horizontal (left panels) and vertical (right panels)
maps of flow velocities at $z=-0.23$, $-0.35$, $-0.47$, and $-0.60$
(top to bottom) Mm for two dark-cored filamentary structures shown in
Fig.\@~2. These flow maps show strong upflows in the dot-like
brightening and weaker {\em downflows} outward and to the sides
of the strong upflow, clearly demonstrating evidence for overturning
convection within the gap. The downflows are not associated
with polarity reversals in Fig. 7 and not aligned with the magnetic field. 
The simulated flows are therefore quite different from (siphon) flows within flux
tubes. Above
$z=-0.35$, the magnitude of the upflow strongly decreases with height whereas the
horizontal outflow peaks at that height. The dark cores are thus
associated with flows that have a strong horizontal flow component near the photoshere
and that are roughly aligned with the dark core. Similar flows along
dark cores were found also in the umbral dot simulations by
Sch\"ussler and V\"ogler (2006). Inspection of movie sequences of
  the horizontal and vertical flow components at $\tau=1$ verifies
  that strong outflows are indeed seen in all gaps. Flows along dark cores were 
  reported by Rimmele and Marino
  (2006) and Langhans et al.\@ (2006). Although the observed flow channels are
  much longer than in the simulations, the simulated sunspot appears to
  develop flows similar to observed Evershed flows. The
  strong upflows at the innermost footpoint reported by Rimmele and
  Marino (2006) are also reproduced by our simulations. These authors
  interpret the observed upflows as evidence in support of
  uni-directional flows along flux tubes, as seen in the simulations
  by Schlichenmaier (1998a,b). The simulations made here show similar
  upflows, but in connection with a more complex flow pattern
  involving overturning convection.
\begin{figure}[t]
  \includegraphics[width=1.00\hsize] {\figdir/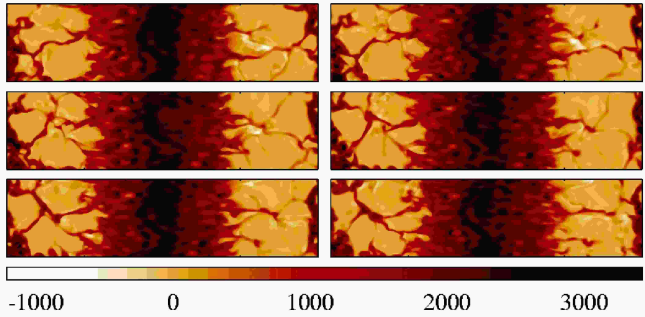}
  \caption{\small Time sequence illustrating the moat flow.  The
    panels (size $12\times3$~Mm) show the vertical magnetic field component at 
    $\tau=1$ at time intervals of 200~s. The frames are ordered
    top-left, top-right, ..., bottom-right.}
  \label{cores7}
\end{figure}

\subsection{Moat flows}

Figure 1 shows an accumulation of magnetic flux at the boundaries of
the computational box and also in the deeper layers below the
atmosphere . This accumulation of flux is in part due pumping down of
high-$\beta$ magnetic gas by convection, and in part due to a
large-scale circulation pattern, shown by arrows in the lower panel.
Movies of the vertical component of the magnetic field at
  $\tau=1$ show a continuous shredding of magnetic flux from the
  sunspot and an outward movement of these structures; a time sequence
  of such a `magnetogram' sequence is shown in Fig.\@~7. We note that
  some of this shredded flux has a polarity that is opposite to that
  of the spot. Our simulation therefore appears to produce moving
magnetic features and moat flows similar to those seen in magnetogram 
and continuum movies (e.g, Hagenaar and Shine 2005).

\section{Discussion and conclusions}

The present simulations, though of restricted scope, already
demonstrate that nearly field-free gaps in the penumbra do form,
as inferred from the observations by SS06a. The presence of up-flows
as well as down-flows within the field-free gaps shows that these
gaps are associated with overturning convection, rather than
uni-directional (siphon) flows. The structures found in the
simulations have several properties in common with observed filaments.
They propagate into the umbral magnetic field, and have a dark core
overlying the center. The inward facing head of the structure
looks similar to tips of filaments, as observed at the boundary
between umbra and penumbra.  The magnetic field {in the simulations}
shows strong fluctuations of the magnetic field inclination across
dark-cored filaments. These variations agree with a number of
independent observations (see discussion in SS06a and references
therein).  The dark-cored structures also are associated with
outflows that appear similar to Evershed flows and
strong upflows at their innermost footpoints, as observed (e.g, Rimmele
and Marino 2006).
Synthetic images calculated for a heliocentric distance of 45$^\circ$
show distinct differences between the limb side and disk center side
that agree with observations. These are the presence of bright dot-like 
features on the limb side but not disk center side (S\"utterlin et al.\@ 2004, 
Tritschler et al. 2004, Langhans et al.\@ 2006) and the better visibility 
of dark cores on the disk center side (S\"utterlin et al.\@ 2004, Langhans 
et al.\@ 2006).

  The simulated sunspot develops a large scale moat flow that
  carries fragments of magnetic flux from the sunspot to the boundary
  of the computational box. Some of this flux shows up with a polarity
  opposite to that of the sunspot. These aspects of the
  simulations are also in qualitative agreement with the
  observations.

In other respects, the structures appearing in the simulations are not
yet fully realistic. The filamentary structures are quite
short, and resemble the structure observed near the umbra/penumbra
boundary more than the penumbra itself. Also, observations indicate
filament lifetimes on the order of 1--2~h, whereas the simulated
filaments form and disappear within a typical time span of 30~min.

These differences may be due to the shallow depth and limited
horizontal extent of the computational box, which in turn restricts
the size of the umbra to a (linear) diameter of approximately 4~Mm.
Indeed, observations occasionally show pores with diameters as large
as 7~Mm without penumbrae (Bray and Loughhead 1964).  An aggravating
factor could be the periodic boundary condition of the simulation in
the `radial' direction, which is equivalent to having umbrae of the
same polarity next to each other. In observations of neighboring spots
of the same polarity, the penumbra is usually suppressed on the side
facing the neighboring spot. The lower boundary condition used, which
keeps the field fixed there, possibly restricts the horizontal extent
of developing structure to something comparable with the depth of the
box.  This could be tested with changes in the lower magnetic boundary
condition or its depth.

Overall, we conclude that the approach to modeling sunspot fine
structure by restricting the computational box to a rectangular
section of a spot is a viable method to deal with the huge range
of scales associated with a mature sunspot.

A number of improvements of these simulations are desirable.
In particular, we will in future simulations
increase the size of the computational box in the direction of the
magnetic field, to allow sections of larger sunspots to be simulated.
The grid separation needs to be reduced in order to adequately resolve
the strong vertical gradients associated with radiative cooling in the
surface layers and also to represent more accurately the small
horizontal scales of the simulated filamentary structures.
  Reduced grid separation is also needed for a more accurate estimate
  of the field strengths in the gaps. In the present simulations, the
  gaps have field strengths on the order of 600--700~G; it is likely
  that this field strength will be reduced with reduced grid
  separation.  Another desirable improvement concerns the
radiative energy transfer; in the present calculations assumed to be
grey and represented by a small number of rays only. A binned opacity
treatment, and an increased number of rays should be employed in
future experiments.

Finally, Doppler and Stokes spectra calculated from the simulation
data will allow `forward modeling' and direct comparisons with
dopplergrams and magnetograms.

\begin{acknowledgements}
  TH thanks Nordita and DAMTP for financial support and Axel Brandenburg and
  Wolfgang Dobler for extensive discussions relating to implementation
  of boundary conditions.  \AN\ thanks JILA and the Danish Natural
  Science Research Council (FNU) for support during a sabbatical stay
  as a JILA Visiting Fellow. Computing time provided by the Danish
  Center for Scientific Computing (DCSC) is gratefully acknowledged.
\end{acknowledgements}

\end{document}